\documentclass[12pt]{article}
\usepackage{a4wide,wrapfig,rotating}
\def\lapproxeq{\lower .7ex\hbox{$\;\stackrel{\textstyle
<}{\sim}\;$}}
\def\gapproxeq{\lower .7ex\hbox{$\;\stackrel{\textstyle
>}{\sim}\;$}}

\def\beq{\begin{equation}}
\def\eeq{\end{equation}}
\def\bea{\begin{eqnarray}}
\def\eea{\end{eqnarray}}

\def\GeV{\rm GeV}
\def\msb{\overline{\rm MS}}

\def\be{\begin{equation}}
\def\ee{\end{equation}}

\begin{document}
\titlepage

\begin{flushright}
LCTS/2012-03\\
\end{flushright}

\vspace*{0.2cm}

\begin{center}
{\Large \bf  Effect of Changes of Variable Flavour Number Scheme on Parton 
Distribution Functions and Predicted Cross Sections}

\vspace*{1cm}
\textup{R.S. Thorne \\
Department Of Physics and Astronomy, University College London\\
          Gower Place, London, WC1E 6BT, UK\\ 
        E-mail: thorne@hep.ucl.ac.uk}
\vspace*{0.2cm} 

\end{center}

\vspace*{0.2cm}

\begin{abstract}
I consider variations in the definitions, at next-to-leading order
(NLO) and at next-to-next-to leading order (NNLO), of a General-Mass 
Variable Flavour Number Scheme (GM-VFNS) for heavy flavour structure 
functions. I also define a new ``optimal'' 
scheme choice improving the smoothness of the transition from one 
flavour number to the next. I investigate 
the variation of the structure function for a fixed set of parton 
distribution functions (PDFs) and also the change in the PDFs when a new 
MSTW2008-type global fit to data is performed for each GM-VFNS. 
At NLO the parton distributions, and predictions using them at hadron 
colliders, can vary by $\sim 2-3\%$ from the mean value. 
At NNLO there is far more stability with varying GM-VFNS 
definition, and changes in 
PDFs and predictions are less than 1$\%$, with most variation at very small
$x$ values. Hence, mass-scheme
variation is an additional and significant source of uncertainty when 
considering parton distributions, but as with all perturbative uncertainties,
it diminishes quickly as higher orders are included.
\end{abstract}

\vspace*{0.2cm}

The treatment of heavy flavours in structure functions has a significant effect on 
the parton distribution functions (PDFs) obtained in fits to structure function, and other 
data, and consequently on the predictions for cross sections at hadron colliders 
such as the LHC and Tevatron. 
The up, down and strange quark are regarded as ``light'' quarks, and are always 
treated in the massless approximation since their current 
mass is much smaller than the regimes in which we use perturbative QCD. However,
both the charm quark mass $m_c$ and bottom quark mass 
$m_b$ are much larger than $\Lambda_{\rm QCD}$ and so we can treat
the generation of the charm and bottom parton distributions as a perturbative 
process. The details of the description of these heavy flavours, charm 
and bottom, in structure functions has an important 
impact on the PDFs extracted in fits, due both 
to the existence of direct data on 
$F_2^h(x,Q^2)$, where $h$ represents the heavy quarks charm $(c)$ 
and bottom $(b)$, and also due to the contribution to the heavy quarks to the 
total structure function, particularly 
at small $x$. Indeed, using the published data \cite{:2009wt}
from the HERA experiment the latter of these is probably more important. 

Traditionally there are two distinct regimes for heavy quark production 
using quite distinct theoretical descriptions.  
For $Q^2\sim m_h^2$ massive quarks are thought of as being
created in the final state, and described using the Fixed Flavour Number 
Scheme (FFNS) (see \cite{Laenen:1992zk} for NLO results), 
\begin{equation}
F(x,Q^2)=C^{\rm FF, n_f}_k(Q^2/m_h^2)\otimes f^{n_f}_k(Q^2),
\end{equation}
where $n_f$ is the number of light quark flavours. 
This does not sum $\alpha_S^n \ln^n Q^2/m_h^2$ terms in the 
perturbative expansion, the consequence of which may well limit accuracy, though
this is still a matter of debate. 
At high scales, $Q^2 \gg m_h^2$, heavy quarks behave like 
massless partons. The large logarithms mentioned can then be automatically 
summed via the solution of evolution equations for the heavy quark parton 
distributions. 
The distributions for different light quark number are related to each other 
via the perturbative expression
\begin{equation}
f^{n_f+1}_j(\mu_F^2)= A_{jk}(\mu_F^2/m_H^2)\otimes f^{n_f}_k(\mu_F^2),
\label{eq:pdfplus1}
\end{equation}
where the matrix elements $A_{jk}(\mu_F^2/m_H^2)$, calculated at 
${\cal O}(\alpha_S^2)$ in \cite{Buza:1996wv} (and where calculations are 
ongoing at ${\cal O}(\alpha_S^3)$ 
\cite{Bierenbaum:2009mv,Ablinger:2010ty,Bierenbaum:2010jp}), 
contain the fixed-order 
$\ln(\mu_F^2/m_h^2)$ contributions. These expressions can be used at an 
appropriate choice of the factorisation scale, usually $\mu_F^2=m_h^2$ to 
obtain the boundary conditions for the evolution of the heavy flavour 
distribution, and simultaneously the change in the light parton distributions 
at this ``transition point''.
In the $Q^2/m_h^2 \to \infty$ limit 
the description becomes the Zero-Mass Variable Flavour Number 
Scheme (ZM-VFNS),
\begin{equation}
F(x,Q^2) = C^{\rm ZMVF,n_f+m}_j\otimes f^{n_f+m}_j(Q^2),
\end{equation}
where $m$ is the number of heavy quarks which have effectively become light 
quarks.
Although called a ``scheme'' this is really an approximation until one reaches 
the real asymptotic limit since it ignores all ${\cal O}(m_h^2/Q^2)$ 
corrections. 

Various studies of comparisons between the FFNS and ZM-VFNS have been made.
For the currently most obvious case of neutral current $F_2^h(x,Q^2)$ it 
seems that the difference at NLO at high $Q^2$ can be significant, but not 
enormous, see e.g. \cite{Gluck:1993dpa,Aivazis:1993pi,Thorne:1997ga}.
However, there are cases where the difference is more obvious, e.g. 
$F_3^h(x,Q^2)$ in charged-current DIS \cite{Buza:1997mg,Thorne:2000zd}, and 
for example there is a distinct difference between the results using the 
two methods 
of calculation for the $b$-quark associated contribution to the $Z$ 
cross-section in hadron-hadron colliders \cite{Martin:2010db} 
and for the neutral supersymmetric 
Higgs boson \cite{Dittmaier:2011ti}, particularly for the largest boson masses. 
It seems natural to assume that the ZM-VFNS is more precise at very large
scales, where resummation of large logarithms is most urgent, and that
the FFNS is more precise for scales near $m_h$, where the ZM-VFNS is only
an approximation. In order to 
correct this shortcoming in the ZM-VFNS but also sum the logarithms
in $Q^2/m_h^2$ (via explicit heavy quark parton distributions), and 
hence to obtain the most effective 
description between the two limits of $Q^2\leq m_H^2$ and 
$Q^2\gg m_H^2$, one must use a General-Mass Variable Flavour Number Scheme 
(GM-VFNS).   

A GM-VFNS is defined similarly to the ZM-VFNS, i.e. the structure function 
is written as 
\begin{equation}
F(x,Q^2) = C^{\rm GMVF,n_f+m}_j(Q^2/m_h^2)\otimes f^{n_f+m}_j(Q^2),
\end{equation}
where now the coefficient functions are dependent on $Q^2/m_h^2$, but 
reduce to the zero mass limit as $Q^2/m_h^2 \to \infty$. If we consider 
the transition from $n_f$ active quarks to $n_f+1$ then we can write
\begin{eqnarray}
F(x,Q^2) \!\!&=& \!\! C^{\rm GMVF,n_f+1}_j(Q^2/m_h^2)\otimes f^{n_f+1}_j(Q^2) \cr 
&=&\!\!
C^{\rm GMVF,n_f+1}_j(Q^2/m_h^2)\otimes A_{jk}(Q^2/m_h^2)\otimes f^{n_f}_k(Q^2) 
\equiv C^{\rm FF, n_f}_k(Q^2/m_h^2)\otimes f^{n_f}_k(Q^2),
\end{eqnarray}
and the GM-VFNS can be defined from the formal equivalence of the 
$n_{f}$ flavour and $n_f+1$ flavour descriptions at all 
orders, resulting in 
\begin{equation}
C^{\rm FF,n_f}_k(Q^2/m_h^2) \equiv 
C^{\rm GMVF,n_f+1}_j(Q^2/m_h^2)\otimes A_{jk}(Q^2/m_h^2),
\label{GMVFNSdeffull}
\end{equation}
where for simplicity I have set $\mu_F^2=Q^2$. The fact that Eq.(\ref{eq:pdfplus1})
correctly converts $n_f$ flavour PDFs to $n_f+1$ flavour PDFs guarantees that 
at $Q^2/m_h^2 \to \infty$, where all power-suppressed $m_h^2/Q^2$ corrections 
become negligible, the GM-VFNS coefficient functions become identical to the 
ZM-VFNS coefficient functions. 

In order to illustrate the manner in which the GM-VFNS works we see 
that at ${\cal O}(\alpha_S)$ Eq.(\ref{GMVFNSdeffull}) 
results in the equivalence 
\be
C^{\rm FF,n_f,(1)}_{2,hg}(Q^2/m_h^2) = 
C^{\rm GMVF,n_f+1,(0)}_{2, h\bar h}(Q^2/m_h^2)\otimes P^0_{qg}\ln(Q^2/m_h^2)+
C^{\rm GMVF,n_f+1,(1)}_{2,hg}(Q^2/m_h^2),
\label{GMVFNSdef1}
\ee
(where $(n)$ represents the $n_{\rm th}$ order in $\alpha_S$ term of the 
coefficient
function), which defines the GM-VFNS coefficient functions.  
As stated, the coefficient functions must tend to the massless limits
as $Q^2/m_h^2 \to \infty$, and Eq.({\ref{GMVFNSdef1}}) is consistent with this.
However, $C^{\rm GMVF}_j(Q^2/m_h^2)$ is only uniquely defined in this limit. 
One can swap ${\cal O}(m_h^2/Q^2)$
terms between $C^{\rm GMVF,(0)}_{2, h \bar h}(Q^2/m_h^2)$ 
and $C^{\rm GMVF(1)}_{2,g}(Q^2/m_h^2)$ in Eq. (\ref{GMVFNSdef1}), 
and make similar swaps at higher order, i.e. as 
pointed out in  \cite{Thorne:1997ga} Eq. (\ref{GMVFNSdeffull}) is just an 
equivalence which can be used in the 
definition of a scheme, but has no unique solution since there are more free
variables on the right-hand side of the equation than there are equations.
(A similar observation was made 
independently in a slightly different context in \cite{Cacciari:1998it}, and
presented very clearly and used to define a particularly efficient scheme
in \cite{Kramer:2000hn}.) To show how the independence under changes of order 
${\cal O}(m_h^2/Q^2)$ works let us consider the expression for 
$F_2^{h\bar h}$ up to ${\cal O}(\alpha_S)$. 
\begin{eqnarray}
F_2^{h\bar h} \!\!&=&\!\! C^{\rm GMVF,n_f+1,(0)}_{2, h\bar h}(Q^2/m_h^2)\otimes (h + \bar h) + \alpha_S C^{\rm GMVF,n_f+1,(1)}_{2, h\bar h}(Q^2/m_h^2,\mu_F^2)\otimes (h + \bar h)\cr 
&+&\!\! \alpha_S \biggl(C^{\rm FF,n_f,(1)}_{2,hg}(Q^2/m_h^2)  
-C^{\rm GMVF,n_f+1,(0)}_{2, h\bar h}(Q^2/m_h^2)\otimes P^0_{qg}\ln(\mu_F^2/m_h^2)\biggr)\otimes g
,
\end{eqnarray}
and let us assume that this is a particular well-defined GM-VFNS. Now we consider making a redefinition
\begin{equation}
C^{\rm GMVF,n_f+1,(0)}_{2, h\bar h}(Q^2/m_h^2) \to C^{\rm GMVF,n_f+1,(0)}_{2, h\bar h}(Q^2/m_h^2)
+ \delta C^{\rm GMVF,n_f+1,(0)}_{2, h\bar h}(Q^2/m_h^2),
\end{equation}
where $\delta C^{\rm GMVF,n_f+1,(0)}_{2, h\bar h}(Q^2/m_h^2)$ vanishes like a power of $m-h^2/Q^2$ as 
$Q^2/m_h^2 \to \infty$. Under such a change in the coefficient function the change in 
the structure function is 
\begin{equation}
\delta F_2^{h\bar h} = \delta C^{\rm GMVF,n_f+1,(0)}_{2, h\bar h}(Q^2/m_h^2)\otimes (h + \bar h)    
 -  \alpha_S \delta C^{\rm GMVF,n_f+1,(0)}_{2, h\bar h}(Q^2/m_h^2)\otimes P^0_{qg}\ln(\mu_F^2/m_h^2)\otimes g.
\label{eq:deltaF}
\end{equation}
From the relationship between the $n_f+1$ and $n_f$ PDFs in Eq.(\ref{eq:pdfplus1}) one can 
easily show that up to ${\cal O} (\alpha_S)$, $\delta F_2^{h\bar h}=0$ (i.e. $(h + \bar h)=
\alpha_S P^0_{qg}\ln(\mu_F^2/m_h^2)\otimes g + {\cal O}(\alpha_S^2)$). However, it is also 
useful to consider the renormalisation group equation in terms of factorisation scale 
$\mu_F$. Clearly a correctly defined structure function will have no dependence on $\mu_F$.   
Explicitly
\begin{eqnarray}
\frac{d\, \delta F_2^{h\bar h}}{d \ln \mu_F^2} \!\!&=& \!\!\delta 
C^{\rm GMVF,n_f+1,(0)}_{2, h\bar h}(Q^2/m_h^2)\otimes \frac{d\,(h + \bar h)}{d \ln \mu_F^2}
- \alpha_S \delta C^{\rm GMVF,n_f+1,(0)}_{2, h\bar h}(Q^2/m_h^2)\otimes P^0_{qg} \otimes g 
+ {\cal O}(\alpha_S^2),\cr
&=& \!\!\alpha_S \delta C^{\rm GMVF,n_f+1,(0)}_{2, h\bar h}(Q^2/m_h^2)\otimes  (P^0_{qg}\otimes g 
 + P^0_{qq} \otimes (h + \bar h)) \cr 
&-& \!\!\alpha_S \delta C^{\rm GMVF,n_f+1,(0)}_{2, h\bar h}(Q^2/m_h^2)
\otimes P^0_{qg} \otimes g + {\cal O}(\alpha_S^2),
\end{eqnarray}
and using the fact that $(h + \bar h))$ starts at ${\cal O}(\alpha_S)$ we see that 
\begin{equation}
\frac{d\, \delta F_2^{h\bar h}}{d \ln \mu_F^2}= 0 + {\cal O}(\alpha_S^2).
\end{equation}   
It can be shown explicitly that this feature exists to higher orders, but it is 
guaranteed from the definition Eq.(\ref{GMVFNSdeffull}) to work at all orders. At a finite
order there is not exact cancellation, but the dependence on the choice of e.g. 
$\delta C^{\rm GMVF,n_f+1,(m)}_{2, h\bar h}(Q^2/m_h^2)$, will be higher order than 
that to which the structure functions are defined. For example, at NLO, where the 
${\cal O}(\alpha_S)$ coefficient functions are applied in the $n_f+1$ flavour regime,
there is no cancellation in Eq.(\ref{eq:deltaF}) between the NLO evolution of the charm 
distribution $\propto \alpha_S^2 P^1_{qg}\otimes g$ in the direct heavy flavour 
contribution and the coefficient function combined with gluon distribution contribution. 
However, this occurs when the ${\cal O}(\alpha_S^2)$ coefficient functions are introduced 
at NNLO, and the uncancelled contribution is pushed to the evolution term 
$\propto \alpha_S^3 P^2_{qg}\otimes g$. As we will see, this results in far more stable 
results under variation of GM-VFNS at NNLO than at NLO. 

Perhaps the nomenclature ``scheme''
is misleading, since it suggests that one compensates for a change in coefficient functions
with a change in PDFs, e.g. as when transforming from the $\msb$ factorisation scheme to
the DIS factorisation scheme. Here it is instead the case that because the heavy flavour is 
generated entirely from the gluon and light quarks via evolution
(ignoring intrinsic flavour, which is discussed in \cite{Thorne:2008xf}) there is an 
inherent ambiguity in exactly which contribution can be attributed to the heavy flavour 
and which to the light PDFs since one can be expressed precisely in terms of the other.
It is only the fact that the large $\ln(Q^2/m_h^2)$ terms are resummed in the heavy flavour 
distribution by evolution that stops the invariance under change of GM-VFNS being exact
at finite order. Expressing 
$(h + \bar h)$ at the same order in $\alpha_S$ as the coefficient functions, rather 
than as the full solution of the evolution equations would lead to exact invariance, but 
would be pointless as it would simply reproduce the results of the FFNS. 

It should be noted that the freedom to modify coefficient functions by power suppressed terms, 
so long as this modification is applied to the corresponding subtraction terms, 
occurs separately in each structure function or cross-section. One may even change the 
GM-VFNS for $F_2(x,Q^2)$ while leaving the one for $F_L(x,Q^2)$ the same. The choice 
made in the fit to structure functions does not influence the choice made in predictions at 
hadron colliders, though at the moment most such predictions are at high enough scales that 
the zero-mass coefficients may be used as a good approximation. This is certainly the case 
for data such as high-$E_T$ jets and $W,Z$ production at the Tevatron and LHC currently 
used in global fits. The heavy flavour contribution to the cross section for 
fixed target Drell Yan data is extremely small so the total is insensitive to 
the heavy flavour prescription used. In principle some collider data would be 
best described using a GM-VFNS. There is no restriction on the form chosen from the manner 
in which the PDFs are fit (unless exactly the same data type is used in the fit, in which 
case there should be consistency). One would simply expect a {\it sensible} definition to 
be used  for all data types. One clear criterion might be that all coefficient functions 
respect the correct kinematics to produce the physical final state containing the heavy quarks.

The freedom to define different definitions of a GM-VFNS 
has resulted in the existence of various prescriptions 
\cite{Aivazis:1993pi,Thorne:1997ga,Chuvakin:1999nx,Tung:2001mv,
Thorne:2006qt} each with a particular reasoning for the choice.
The original ACOT scheme \cite{Aivazis:1993pi} simply calculated the 
heavy quark coefficient functions for single massive quark production and 
defined these to be $C^{\rm GMVF,n_f+1,}_{2, h\bar h}$, hence defining the scheme.
However, this required explicit calculation and also did not impose 
the correct kinematical requirement that (in neutral current DIS) one
must have enough energy to create a pair of massive quarks in the final state. 
The TR GM-VFNS \cite{Thorne:1997ga} highlighted the freedom in choice, 
and enforced correct kinematics via a  definition
which demanded the continuity of the value of $dF_2^h/d \ln Q^2$
across the transition point. However, this was only possible in the gluon sector 
beyond ${\cal O}(\alpha_S)$ and was quite complicated. 
The (S)ACOT($\chi$) prescription \cite{Tung:2001mv} applied the simple choice 
\begin{equation}
C^{\rm GMVF,(0)}_{2, h \bar h}(Q^2/m_h^2,z)\propto \delta(z-x_{\max}),
\end{equation}
which gives 
\be
F^{h,(0)}_2(x,Q^2)\propto e_h^2 (h+\bar h)(x/x_{\max}, Q^2), 
\ee
where  $x_{\max}=Q^2/(Q^2+4m_h^2)$,
and imposes the threshold $W^2=Q^2(1-x)/x \ge 4m_h^2$. 
This gives the usual limit 
\be
C^{\rm ZMVF,(0)}_{2, h \bar h}(z)= \delta(1-z)
\ee
for $Q^2/m_h^2 \to \infty$. The modified TR, or TR' scheme \cite{Thorne:2006qt}
adopted this and extensions to higher orders (though uses a 
different multiplicative factor of $Q^2/(Q^2+4m_h^2)$ 
for the zeroth-order coefficient function \cite{Forte:2010ta}). 
Very recently the (S)ACOT($\chi$) prescription has been extended explicitly to 
NNLO \cite{Guzzi:2011ew}.

However, there is a further, more significant difference between the 
TR' scheme and the SACOT$(\chi)$ scheme. ACOT-type schemes have used 
the same order of $\alpha_S$ above and below the transition point 
$Q^2=m_h^2$, despite
the fact that FFNS is LO at ${\cal O}(\alpha_S)$ while the ZM-VFNS starts at 
zeroth order. Instead the TR' definition uses, for example, at LO 
the ${\cal O}(\alpha_S)$ FFNS result for $Q^2<m_h^2$, and for
$Q^2>m_h^2$
\be
F_2^h(x,Q^2)=\alpha_S(m_h^2) 
C^{\rm FF,n_f,(1)}_{2,hg}(Q^2=m_h^2)\otimes 
g^{n_f}(Q^2=m_h^2)
+ C^{\rm GMVF,n_f+1,(0)}_{2, h \bar h}(Q^2/m_h^2)
\otimes (h+\bar h)(Q^2),
\ee
i.e. it freezes the higher order $\alpha_S$ term when going upwards 
through $Q^2=m_h^2$.
This difference in choice can be phenomenologically important in the vicinity of
$Q^2=m_h^2$, though becomes less important at high $Q^2$ as the frozen term 
becomes smaller compared to the growing contribution from the heavy quark 
distribution. The effect is quite significant at small $x$ at NLO, since the 
${\cal O}(\alpha_S^2)$ contribution $\alpha_S^2 C^{\rm FF,n_f,(2)}_{2,cg}\otimes 
g^{n_f}$ is large for $Q^2 \sim m_h^2$, as seen in Fig. 2 of 
\cite{Thorne:2006qt}.    

There is also an alternative, but ultimately equivalent, 
way of formulating a GM-VFNS, where instead of having a transition point where
one turns on heavy quark distributions and defining GM-VFNS coefficient functions
one obtains the total structure function from a variety of contributions which
cancel each other appropriately to obtain the correct limits in the low and
high $Q^2$ regimes. The BMSN (Buza, Matiounine, Smith and van Neerven) 
\cite{Buza:1996wv} and FONLL (fixed-order next-to-leading log) 
\cite{Forte:2010ta} definitions for a scheme are based on the expression  
\be
F^{\rm GMVF}(x,Q^2) = F_2^{\rm FF}(x,Q^2) -
F_2^{\rm asymp}(x,Q^2)+F_2^{\rm ZMVF}(x,Q^2),
\ee
where generally speaking the second (subtraction) term is the 
asymptotic version of the first, 
i.e. all terms of ${\cal O}(m_h^2/Q^2)$ are omitted.
There are differences in exactly how the second and third terms are defined 
in detail in different schemes. 
In the standard applications the $\alpha_S$ order of 
$F^{\rm FF}_2(x,Q^2)$ at low $Q^2$ is the same as that of  
$F^{\rm ZMVF}_2(x,Q^2)$ as $Q^2\to \infty$. This type of scheme is used 
in \cite{Alekhin:2009ni} using ${\cal O}(\alpha_S)$ coefficient function
expressions along with LO PDFs, and ${\cal O}(\alpha_S^2)$ coefficient function
expressions along with NLO PDFs. However, ultimately the FFNS scheme is used 
for structure functions in fitting to structure function data,  
${\cal O}(\alpha_S^2)$ for both NLO and NNLO fits 
(approximate ${\cal O}(\alpha_S^3)$ coefficient functions, and a running 
quark mass, are used in 
\cite{Alekhin:2010sv} and \cite{Alekhin:2012ig}). 
The FONLL approach, which modifies $F_2^{\rm asymp}(x,Q^2)$ and 
$F_2^{\rm ZMVF}(x,Q^2)$ by, for example, the replacement of $x$ by 
$\chi$ in order to maintain correct kinematics in each term, is used at 
NLO in \cite{Ball:2011mu} and at LO and NNLO in \cite{Ball:2011uy}.
By default the order of $\alpha_S$ is the same in each of the three
terms and the GM-VFNS uses ${\cal O}(\alpha^0_S)$ coefficient function
expressions along with LO PDFs, ${\cal O}(\alpha_S)$ coefficient function
expressions along with NLO PDFs and ${\cal O}(\alpha_S^2)$ coefficient function
expressions along with NNLO PDFs.
There is a version of 
FONLL , FONLL-B, which uses one power higher in the FFNS and subtraction
term, but where the subtraction term contains only the $\ln(Q^2/m_h^2)$ terms, 
not the finite part, so that at $Q^2=m_h^2$, 
$F^{\rm GMVF}(x,Q^2) \equiv F_2^{\rm FF}(x,Q^2)$. 
This reproduces  the correct NLO FFNS at $Q^2$ near to threshold, 
but unlike the case where the coefficient functions are ${\cal O}(\alpha_S)$ in 
all pieces it leads to 
part (but not all) of the higher order NNLO contribution persisting as 
$Q^2 \to \infty$, i.e. $F_2^{\rm FF}(x,Q^2)$  and 
$F_2^{\rm asymp}(x,Q^2)$ do not cancel in this limit so the exact
$F_2^{\rm ZMVF}(x,Q^2)$ is accompanied by a formally subleading term, reminiscent
(but different in origin and form) of the frozen contribution in the TR' approach.

Ideally one  
would like any GM-VFNS to reduce to exactly the correct order 
FFNS at low $Q^2$ and exactly the correct order (one power of $\alpha_S$
lower) ZM-VFNS as 
$Q^2 \to \infty$. At present, as outlined above, none do. However, in the case 
of the TR' scheme this can easily be rectified. The fact that the 
higher-$\alpha_S$ order term is frozen for $Q^2>m_h^2$ was strictly
necessary in the original TR scheme in order to impose exact continuity 
of $dF^h_2/d \ln Q^2$. However, this is no longer a requirement, so rather than
this term being frozen it could instead be of the form
\be
(m_h^2/Q^2)^{a} \alpha_S^n(m_h^2)\!\sum 
C_{2,i}^{\rm FF}(m_h^2)\!\otimes \!f_i(m_h^2)  
\qquad {\rm or}\qquad  (m_h^2/Q^2)^{a}\alpha_S^n(Q^2)\!\sum 
C_{2,i}^{\rm FF}(Q^2)\!\otimes\! f_i(Q^2).
\label{parama}
\ee
Any $a >0$ provides the correct asymptotic limit, though strictly from 
factorization one should 
have $(m_H^2/Q^2)$ times $\ln(Q^2/m_H^2)$ terms. Fractional values of
$a$ could be thought of as mimicking the power times logarithmic corrections. 
As well as modifying the previously frozen term in this manner
there is also more freedom in the definition of the GM-VFNS than previously 
used or explored. Indeed, this freedom is an inherent feature of a GM-VFNS,
so its consequences should be investigated quantitatively.

One can modify the heavy quark coefficient 
function so long as the correct $Q^2/m_h^2 \to \infty$ limit is maintained. 
However, since this appears in convolutions for higher order subtraction 
terms, it is desirable to avoid a complicated $x$ dependence. 
A simple choice is 
\be
C^{\rm GMVF,(0)}_{2,h \bar h}(Q^2/m_h^2,z)\to  (1+b(m_h^2/Q^2)^c)\delta(z-x_{\max}),
\label{parambc}
\ee
where again variation in $c$ really mimics $(m_h^2/Q^2)$ with 
logarithmic corrections.  One can also modify the argument 
of the $\delta$-function, similar to the Intermediate-Mass 
IM scheme \cite{Nadolsky:2009ge}, we can define 
\be
\xi = x/x_{\max} \to x \bigl(1+(x(1+4m_h^2/Q^2))^d 4m_h^2/Q^2\bigr),
\label{paramd}
\ee
so the kinematic limit stays the same, but if $d>0 \,(<0)$ small $x$ is less 
(more) suppressed. The default $a,b,c,d$ are all zero, but can vary, being 
limited by fit quality or {\it sensible} choices, e.g. we would not choose
$b < -1$, since this would give a negative zeroth-order
heavy quark coefficient function near $Q^2=m_h^2$. There is also a potential 
source of uncertainty from variation of the transition scale away from 
$\mu^2=m_h^2$. However, much of the type of possible variation in $F_2^h$ is 
very similar to that  
obtained by the variation of parameters already described.

\begin{table}
\begin{center}
\begin{tabular}{|l|l|l|l|l|}
\hline
scheme & a & b & c & d \\
\hline
GM-VFNS1& 0 & -1 & 1 & 0 \\
GM-VFNS2& 0 & -1 & 0.5 & 0 \\
GM-VFNS3& 1 & 0 & 0 & 0 \\
GM-VFNS4& 0 & 0.3 & 1 & 0 \\
GM-VFNS5& 0 & 0 &0 & 0.1 \\
GM-VFNS6& 0 & 0 & 0 & -0.2 \\
optimal& 1 & -2/3 & 1 & 0 \\
\hline
\end{tabular}
\end{center}
\vspace{-0.3cm}
\caption{The values of parameters defined in Eqs.({\ref{parama},\ref{parambc},\ref{paramd}})   for different extreme GM-VFNS definitions.}
\label{tab:GMVFNSdef}
\end{table}

\begin{figure}
\vspace{-0.52cm}
\centerline{\hspace{-1.6cm}\includegraphics[width=0.45\textwidth]{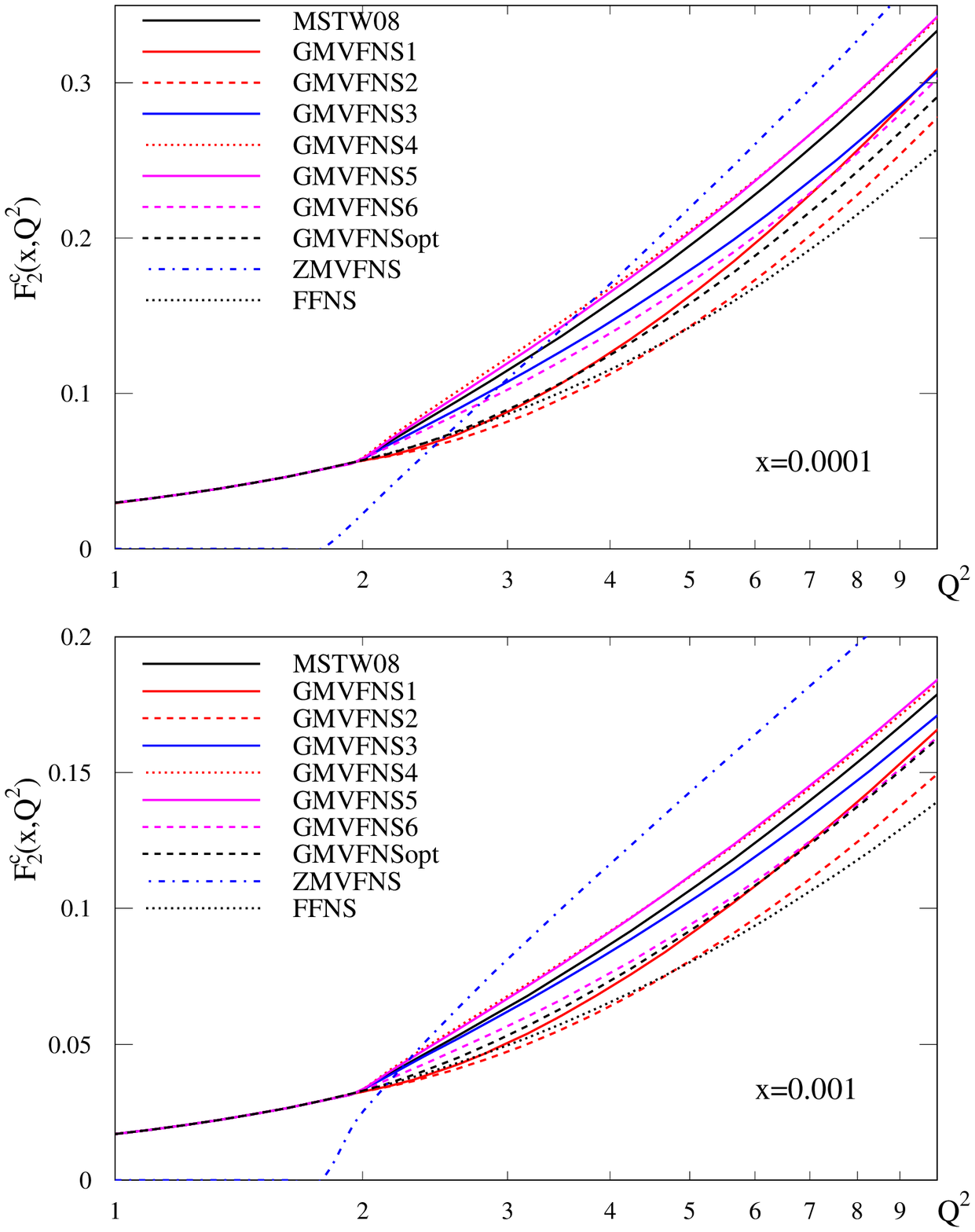}
\includegraphics[width=0.45\textwidth]{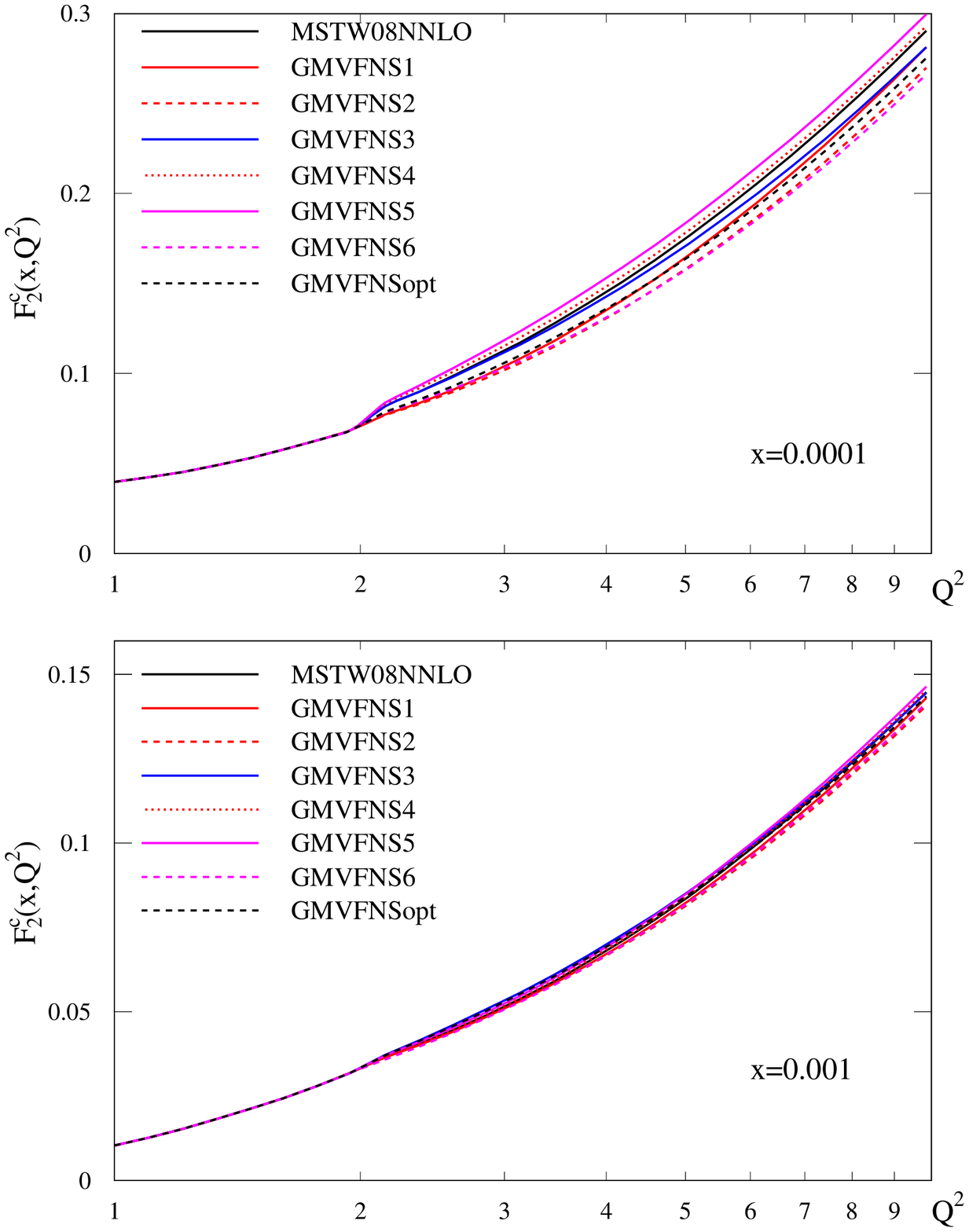}}
\vspace{-1.5cm}
\caption{The variation in $F_2^c(x,Q^2)$ generated from a variety of choices of 
GM-VFNS at NLO (left) and NNLO (right) using the MSTW2008 pdfs in each case.}
\vspace{-0.2cm}
\label{gmvarf2c} 
\end{figure}

\begin{figure}
\vspace{-0.2cm}
\centerline{\hspace{-1.6cm}\includegraphics[width=0.45\textwidth]{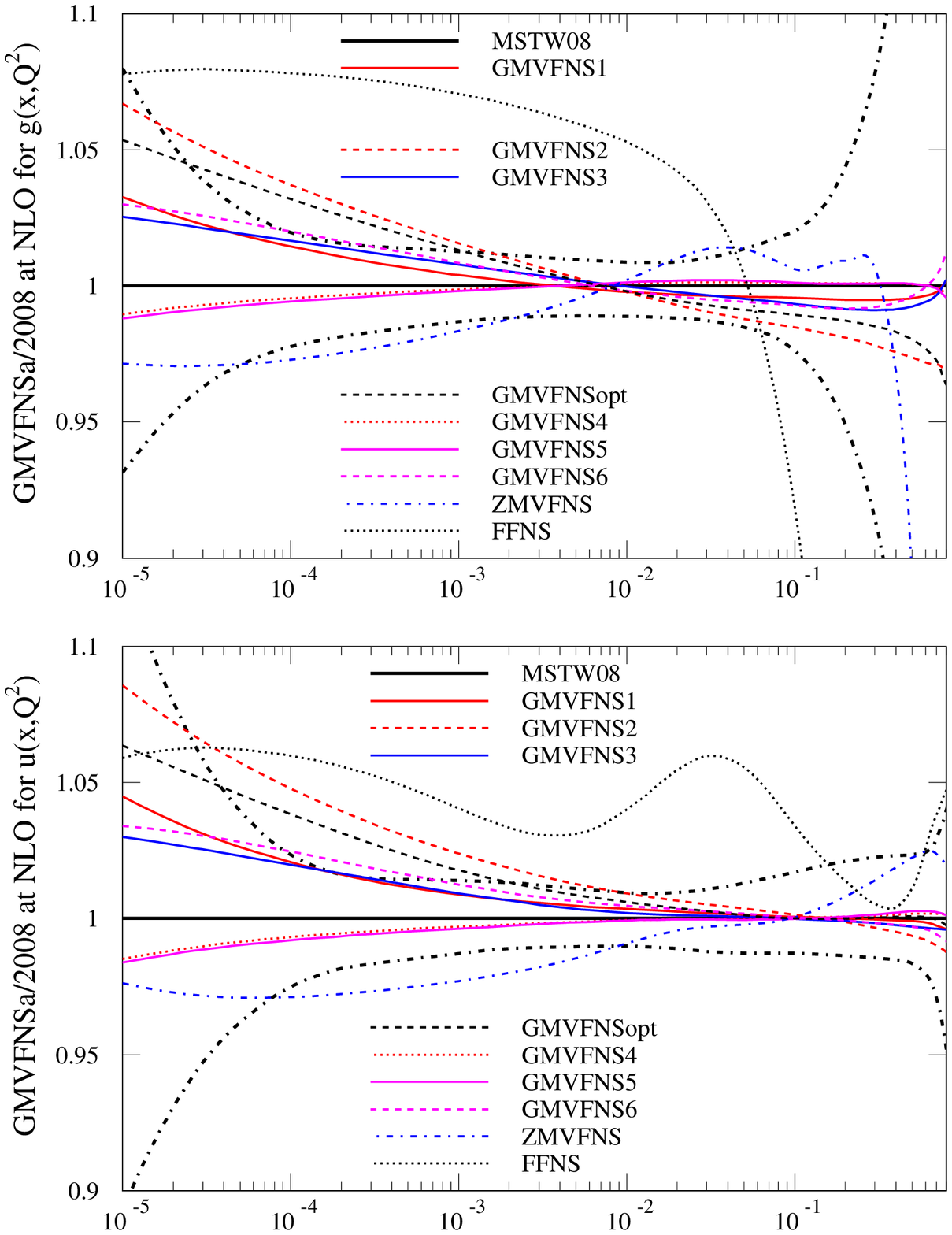}
\includegraphics[width=0.45\textwidth]{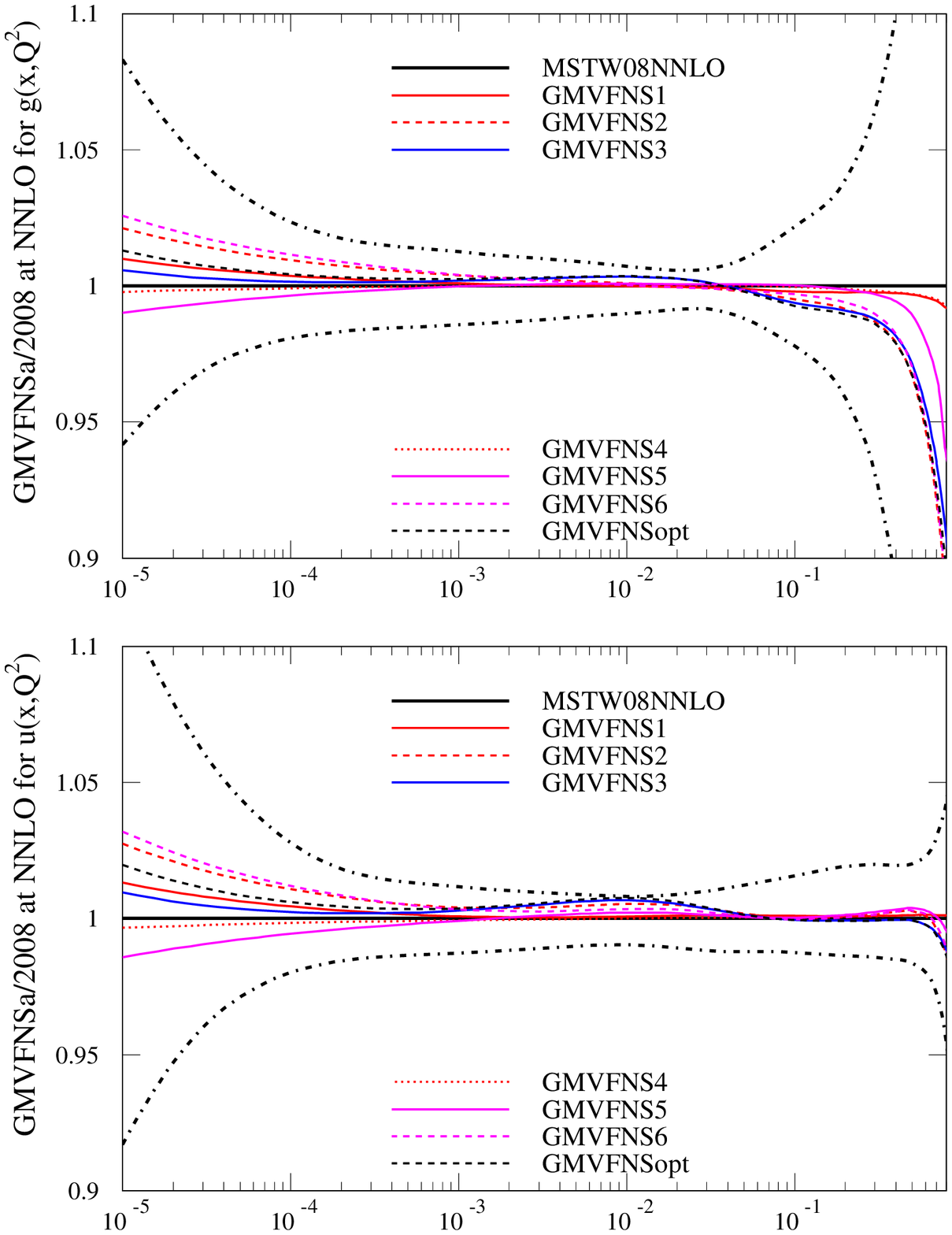}}
\vspace{-1.5cm}
\caption{The variation of the gluon (upper plots) and quark (lower plots) PDFs at $Q^2=10,000\GeV^2$ 
obtained from the best fit from a variety of choices of 
GM-VFNS, the ZM-VFNS and the FFNS at NLO (left) and the GM-VFNS variations at NNLO (right) as a ratio to the MSTW2008 PDFs. The one-sigma uncertainties for the MSTW2008 PDFs are shown as dash-dotted lines.}
\vspace{-0.5cm}
\label{gmvarpart} 
\end{figure}

A variety of different choices defined in Table \ref{tab:GMVFNSdef}
has been tried at NLO and at NNLO, along with the ZM-VFNS (at NLO).\footnote{Some
of the results from this were presented in \cite{Thorne:2010pa}.} 
The resulting variations 
in $F_2^c(x,Q^2)$ near the transition point 
due to different choices of GM-VFNS at NLO are shown in the left of 
Fig. \ref{gmvarf2c}. (A similar variation under changes in scales and scaling 
variable has appeared in \cite{Guzzi:2011ew}.)
I also define an ``optimal'' scheme which is 
chosen to be smooth at threshold, at NLO, 
and reduces to exactly the right limits at high and low $Q^2$. 
There is quite a spread in the values of $F_2^h(x,Q^2)$ at NLO, though the 
ZM-VFNS is far steeper at low $Q^2$ than any GM-VFNS. The continuation of 
the FFNS result above $Q^2=m_h^2$ is also shown at NLO, and it is clear that at 
both $x$ values all GM-VFNS variations are ultimately rising above this. In the vicinity 
of $Q^2=m_h^2$ the optimal scheme stays closest to FFNS. 
The spread is  
very much reduced at NNLO, the right of Fig. \ref{gmvarf2c},
with almost zero variation until very small $x$. At this point the 
precise cancellation between different terms in the expression for the GM-VFNS
near the transition point is starting to fail to some extent,
showing that NNLO evolution effects are most important in this 
regime and even higher orders in the GM-VFNS are needed for there to be 
very little sensitivity to choices. The FFNS is not shown in this case as 
the  NNLO (${\cal O}(\alpha_S^3)$) contribution to this is only approximate
and it is possible, and arguably likely, that the approximation is worse 
away from the $Q^2 \sim m_h^2$ regime where it is important in the GM-VFNS.

\begin{table}
\begin{center}
\vspace{-0.0cm}
\hspace{-0.1cm}
\begin{tabular}{|l|ll|ll|ll|}
\hline
PDF set & Tevatron & & LHC &(${\rm 7~TeV}$)& LHC &(${\rm 14~TeV})$\\
& $\sigma_Z\,{\rm (nb)}$  & $\sigma_H$(pb) 
& $\sigma_Z\,{\rm (nb)}$  & $\sigma_H$(pb) 
& $\sigma_Z\,{\rm (nb)}$ & $\sigma_H$(pb) \\
\hline
       ${\rm MSTW2008}$ & 7.207 & 0.7462 & 27.70 & 12.41 & 59.25 & 40.69  \\   
\hline
       ${\rm GMvar1}$ &    $+0.3\%$ & $-0.5\%$ &  $+0.7\%$ & $-0.1\%$&  $+1.1\%$ & $+0.2\%$   \\    
       ${\rm GMvar2}$ &  $+0.7\%$ &  $-1.1\%$  &  $+2.0\%$ & $+0.5\%$&  $+3.0\%$ & $+1.5\%$   \\    
       ${\rm GMvar3}$ &  $+0.1\%$ &  $-0.3\%$  &  $+0.7\%$ & $+0.4\%$&  $+1.1\%$ & $+0.8\%$  \\    
       ${\rm GMvar4}$ &  $+0.0\%$ &  $-0.1\%$  & $-0.3\%$ & $-0.1\%$ & $-0.4\%$ & $-0.2\%$   \\    
       ${\rm GMvar5}$ & $-0.1\%$ &  $-0.1\%$   & $-0.4\%$ & $-0.2\%$ & $-0.5\%$ & $-0.3\%$   \\    
       ${\rm GMvar6}$ &  $+0.3\%$ &  $-0.4\%$  &  $+1.0\%$ & $+0.3\%$&  $+1.6\%$ & $+0.8\%$   \\
       ${\rm GMvaropt} $& $+0.3\%$ &  $-1.5\%$ &  $+1.5\%$ & $+0.1\%$&  $+2.0\%$ & $+0.4\%$  \\
${\rm Z}$M-V${\rm FNS} $ &$-0.7\%$ &  $-1.2\%$ & $-1.6\%$ & $-1.8\%$ & $-3.0\%$ & $-3.1\%$   \\
  FFNS - DIS only &$+3.1\%$ &  $-14.4\%$ & $+4.5\%$ & $-0.7\%$ & $+7.0\%$ & $+3.0\%$   \\
     ${\rm GMvarcc}$  & $+0.0\%$ &  $-0.1\%$   &  $+0.0\%$ & $-0.1\%$&  $+0.0\%$ & $-0.1\%$  \\    
\hline
    \end{tabular}
\end{center}
\vspace{-0.3cm}
\caption{Predicted cross-sections 
at NLO for 
$Z$ and a 120 GeV Higgs boson at the Tevatron 
and LHC.} 
\label{cstablenlo}
\end{table}

Global fits are also performed using the same procedure as the MSTW2008 fit 
\cite{Martin:2009iq} for all
schemes, and the value of $\alpha_S(M_Z^2)$ is allowed to vary in all cases.
At NLO the initial $\chi^2$ for a new GM-VFNS can change by up to $250$, 
but converges to within $20$ of the original with refitting. 
The fit is improved for 
options 1, 2 and 6, and the optimal scheme and the fit is best   
for the option 6, where it is 23 units lower
than for the best fit in 
\cite{Martin:2009iq} -- in particular the quality of fit to the HERA 
$F_2^c$ data improves from 110 to 90 for 83 points. 
For option 2, the fit quality is almost unchanged. For options 4 and 5 the 
values of $b$ and $d$ respectively are limited by the quality of fit
of the HERA $F_2^c$ data, i.e. we stop variations when the fit quality 
reaches 7 units higher than the best fit for MSTW2008.  
The variations in the partons extracted at 
NLO are shown in the left of Fig. \ref{gmvarpart}. The default TR' scheme 
sits near the low end at small $x$ values.
Some changes in PDFs exceed the one $\sigma$ experimental PDF uncertainty in the 
upwards direction at small $x$. $\alpha_S(M_Z^2)$ changes by $< 0.0007$ 
for all variations of the GM-VFNS, only about half the experimental uncertainty 
found in \cite{Martin:2009bu}.\footnote{A comparison between GM-VFNS definitions of different groups 
using a common set of PDFs can be found in \cite{Binoth:2010ra}, and more 
complete comparisons are in progress. The differences are well-understood. At 
low $Q^2$ some of the differences are bigger than those exhibited in this 
article, but this is due to the different ordering chosen by different groups 
in the $Q^2<m_h^2$ region which is a fixed part of our definition.}
I also perform a fit using the ZM-VFNS. 
In this case the value of $\alpha_S(M_Z^2)$  falls by 
0.0015.  The ZM-VFNS PDF is clearly outside the GM-VFNS band in some 
places.The fit quality is about 200 units worse than the MSTW2008 fit. 
130 of this is the fit to the $F_2^c(x,Q^2)$ data which is very poorly fit at 
low $Q^2$ and much of the rest is from a worse fit to HERA structure function 
data. There is little difference in the fit if the charm structure function 
data are omitted. 

It is also interesting to investigate a fit in 
the FFNS scheme. In this case it is not possible to perform a full
global fit since the coefficient functions are not fully known for hadron-hadron
collider processes even at NLO. It would be possible to perform a hybrid fit
where for structure functions we stayed with the three flavour description but 
for collider data let the same partons evolve in a variable flavour scheme. However, 
this seems a peculiar compromise - either one attempts to resum the logarithms or 
one does not. Hence, I consider a FFNS fit to structure function data only, also 
omitting the high-$Q^2$ charged current HERA data, which also rely on not fully
known coefficient functions at order $\alpha_S^2$. The resulting fit has 
$\chi^2=1944/2089$, compared to $\chi^2=1878/2089$ for the MSTW2008 fit, i.e. the 
contribution of the the $\chi^2$ from these data to the full MSTW2008 fit, not 
a new fit to the restricted data set. A new fit using the default GM-VFNS gives 
an improvement of about 30 units and only minor changes in PDFs. 
The improvement is mainly due to some reshuffling of the
flavours with no Drell Yan data constraint allowing a better fit to the fixed
target DIS data. There is little change for HERA data. 
The fit using the FFNS has a value of $\alpha_S(M_Z^2)$ of 0.1187 when it is 
evolved from low scales in a variable flavour scheme, 0.0025 lower 
than the MSTW value. This lower value allows a better fit to BCDMS data, but there 
is a deterioration in the fit to NMC data, to the nuclear target neutrino DIS data 
and the the HERA data at higher $Q^2$ where $F_2(x,Q^2)$ does not rise as quickly with
$Q^2$ as in the GM-VFNS. The fit to $F^c_2(x,Q^2)$, for which the most precise data 
is at relatively low $Q^2$, is slightly better in the FFNS, but this is far from 
compensating for the deteriorations elsewhere. The partons obtained in the FFNS 
fit are qualitatively different to those in the GM-VFNS fit. The gluon distribution 
is larger at small-$x$ and correspondingly lower at high-$x$. This is qualitatively
very similar to the difference between the gluon distribution in the ABKM PDF set
\cite{Alekhin:2009ni} (or more recently \cite{Alekhin:2012ig}), which uses the FFNS 
and the MSTW set, as is the change in 
the coupling constant. At small-$x$ this difference in the gluon shape is reflected in 
the quarks at higher $Q^2$ since their evolution is determined mainly by the gluon. 
The input PDFs in the FFNS fit can be evolved using a variable flavour scheme in order
to make a more direct comparison at high-$Q^2$ possible, and to examine the 
compatibility with collider data. The ratio to the MSTW2008 PDFs can be seen in 
the left of Fig. \ref{gmvarpart}, and it is clear they are qualitatively different. 
The predictions for Tevatron jet data are very poor, and there is a distinct deterioration 
in the fit quality for the $Z$ rapidity data from CDF. In contrast the DIS-only fit using the
GM-VFNS gives predictions for Tevatron jet data that have $\chi^2$ values only a few units 
higher than when these data are included, and the prediction for the $Z$ rapidity data is 
actually marginally better than the global fit. Clearly at NLO the results from the FFNS
and any GM-VFNS are rather different. One would expect this difference to
diminish at higher orders, but a precise comparison awaits the full calculation of FFNS
coefficient functions.


The variation in GM-VFNS is also applied to the charged-current 
cross-section expressions.  None of the variations in $a,b,c,d$ 
of the form already considered lead
to changes in the $\chi^2$ of much greater than one unit even before 
refitting. The change in the PDFs with refitting is largest for the strange 
quark, which is probed by dimuon production which relies on the heavy flavor
formalism,  
but is always very much smaller than 1$\%$. This is 
due to the fact that the changes in structure function with change in 
scheme definition are by far most significant at small $x$, and the charged 
current data is all for $x > 0.01$. It is also because in the charged-current 
case one only produces a single heavy particle in the final state, i.e. the 
threshold $W^2=m_h^2$ rather than $4m_h^2$ so there
is less sensitivity to how the massless limit is approached than 
the neutral current case. Finally, the data on dimuon production is a 
combination of the structure functions $F_2(x,Q^2)$, $F_L(x,Q^2)$ and 
$F_3(x,Q^2)$, each with a particular $y$ dependence. In practice the 
contribution from the charm quark contribution and the corresponding gluon 
subtraction term are suppressed by a factor of $(1-y)^2$. For these data 
$0.3 < y <0.7$ so this factor is always significantly less than 1, and moreover 
the contribution with larger charm quark or subtraction are at larger $y$ and 
most suppressed. \footnote{There is the related issue of how
diagrams where the final state charm quark is generated away from the parton-boson interaction vertex should be considered in dimuon production, 
which is done differently by different groups and related to the acceptance 
corrections. This should be clarified, but is a separate topic 
to the particular scheme variations considered in this article. 
The suppression noted above makes the effect very small, however.}.

The predictions for cross-sections are shown 
at NLO in Table. \ref{cstablenlo}. It is assumed that any
${\cal O} (m_{c,b}^2/m_{Z,H}^2)$ corrections which would occur in a 
full GM-VFNS definition are negligible and the ZM-VFNS result is used. 
There is at most a $1.5\%$ variation at the Tevatron.  
At the recent energy at the LHC of 7~TeV there is a variation of 
$+2\%$ to $-0.4\%$ in $\sigma_Z$.
This is increased to  $+3\%$ to $-0.5\%$  
at 14~TeV due to smaller $x$ being sampled. The spread in $\sigma_H$ 
at the LHC is about halved compared to $\sigma_Z$ due to 
the higher average $x$ sampled. The ZM-VFNS is a clear outlier 
in the low direction at the LHC. However, the ZM-VFNS variations are only 
about $3\%$ at most compared to our default, and never more than 
$5\%$ even compared to the most upward GM-VFNS variations. Hence, the ZM-VFNS
change is more similar to that observed by NNPDF when adopting a GM-VFNS
to obtain PDFs \cite{Ball:2011mu}, than the larger changes observed by 
CTEQ when adopting a GM-VFNS as default for the first time 
\cite{Tung:2006tb}. The FFNS predictions are also shown, and display some 
rather large variations, particularly in the Tevatron Higgs cross section. 
However, it should be remembered that this is a fit to a more limited data 
set than all the others. It is, however, indicative of the major differences 
seen between MSTW2008 PDFs and some PDFs obtained using fits using the FFNS. 
GMvarcc denotes variation in the GM-VFNS for  
charged current processes, and clearly the effect is very small indeed.

For fits at NNLO the initial changes in $\chi^2$ are much smaller than 
at NLO, i.e. $< 20$, and
they converge to within $10$ of the original.
In fact they are all within 2-3 units of the default fit except for 
options 2 and 6 where the fit quality is about 10 worse. In fact for 
option 6 the variation of the $d$ parameter is limited by the 
deterioration in the fit to the HERA $F_2^c$ data in the NNLO fit
rather than NLO. 
The variations in the PDFs extracted at NNLO are shown in the right of
Fig. \ref{gmvarpart}. At the very 
worst the changes approach the one $\sigma$ experimental PDF uncertainty,
but are usually far less. 
Variations in $\alpha_S(M_Z^2)$ are $\sim 0.0003$. Hence, the stability 
of the PDFs to variations in the parameters $a, b, c, d$ defining how 
quickly the heavy flavour PDF turns on, and correspondingly the size
of subtraction terms, diminishes greatly at NNLO due to increased
cancellation of the two types of term.  

\begin{figure}
\vspace{-0.52cm}
\centerline{\hspace{-1.5cm}
\includegraphics[width=0.6\textwidth]{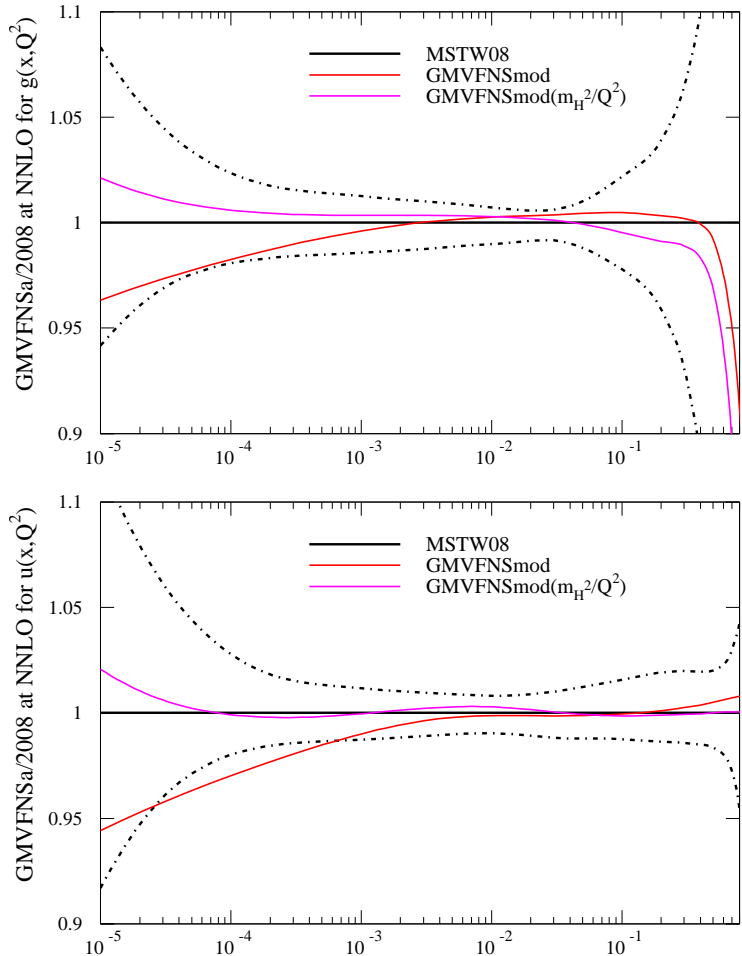}}
\vspace{-1.5cm}
\caption{The variation in PDFs at $Q^2=10,000\GeV^2$ obtained from two choices
of GM-VFNS at NNLO where the coefficient of the unknown constant in the 
small-$x$ term in the ${\cal O}(\alpha_S^3)$ part of the FFNS coefficient 
function is taken to its extreme value. In one case it is frozen for 
$Q^2>m_c^2$ and in the other it falls like $m_c^2/Q^2$.}
\vspace{-0.5cm}
\label{gmvarpartnnlomod} 
\end{figure}

However, at NNLO there is another source of uncertainty associated with 
the GM-VFNS definition. The TR' scheme models the ${\cal O}(\alpha_S^3)$ 
FFNS terms at low $Q^2$ using known leading threshold 
logarithms \cite{Laenen:1998kp} and $\ln(1/x)$ terms \cite{Catani:1990eg}.
The small-$x$ model takes the form
\be
A(Q^2/m_h^2)(1-z/x_{\max})^{\tilde a}(\ln(1/z)-{\tilde b})/z,
\label{sxmodel}
\ee
where $A(Q^2/m_h^2)$ is a calculated function, 
but $\tilde a$ and $\tilde b$ are free parameters determining at what $x$ 
the leading $\ln(1/x)$ contribution becomes dominant. The default values 
are ${\tilde a}=20, {\tilde b}=4$, but ${\tilde a}$ 
and ${\tilde b}$ can be varied. 
Changes in ${\tilde a}$ within reasonable limits 
make little difference. The maximum 
{\it sensible} variation  of ${\tilde b}=2$ 
(determined by the fact that the fit quality is 
not significantly altered, similar to the variations of $a$-$d$) 
leads to an effect 
of order the experimental PDF uncertainty at  $x\leq 0.001$, as shown in 
Fig. \ref{gmvarpartnnlomod}.
However, this change in the PDFs is largely eliminated if the 
${\cal O}(\alpha_S^3)$
contribution dies away like $m_h^2/Q^2$, rather than being frozen, as 
also seen in Fig. \ref{gmvarpartnnlomod}.
This is another reason for preferring this choice.    

\begin{table}
\begin{center}
\hspace{-0.6cm}
\begin{tabular}{|l|ll|ll|ll|}
\hline
PDF set & Tevatron & & LHC &(${\rm 7~TeV})$& LHC &(${\rm 14~TeV})$\\
& $\sigma_Z\,{\rm (nb)}$  & $\sigma_H$(pb) & $\sigma_Z\,{\rm (nb)}$  & $\sigma_H$(pb) 
& $\sigma_Z\,{\rm (nb)}$ & $\sigma_H$(pb) \\
\hline
     ${\rm MSTW2008}$    & 7.448     & 0.9550    & 28.53    & 15.71 & 60.93 & 50.51  \\    
\hline
     ${\rm GMvar1}$    & $+0.1\%$  & $-0.5\%$  & $+0.1\%$  & $-0.3\%$& $+0.1\%$  & $-0.2\%$ \\    
     ${\rm GMvar2}$    & $+0.3\%$  & $-0.8\%$  & $+0.2\%$  & $+0.1\%$& $+0.5\%$  & $+0.1\%$\\    
     ${\rm GMvar3}$    & $+0.4\%$  & $-0.1\%$  & $+0.4\%$  & $+0.6\%$& $+0.5\%$  & $+0.7\%$ \\    
     ${\rm GMvar4}$    & $+0.0\%$  & $-0.2\%$  & $-0.1\%$  & $-0.1\%$& $+0.1\%$  & $-0.1\%$\\    
     ${\rm GMvar5}$    & $+0.1\%$  & $-0.3\%$  & $-0.1\%$  & $-0.2\%$& $-0.2\%$  & $-0.2\%$\\    
     ${\rm GMvar6}$    & $+0.1\%$  & $-0.9\%$  & $+0.0\%$  & $-0.5\%$& $+0.3\%$  & $-0.2\%$\\
     ${\rm GMvaropt}$  & $+0.4\%$  & $-0.2\%$  & $+0.5\%$  & $+0.6\%$& $+0.6\%$  & $+0.8\%$ \\
     ${\rm GMvarmod }$  & $-0.2\%$ & $-0.4\%$  & $-0.8\%$  & $-0.6\%$& $-1.4\%$  & $-1.0\%$ \\
     ${\rm GMvarmod'}$ & $+0.0\%$  & $-0.7\%$  & $+0.0\%$  & $+0.0\%$& $+0.0\%$  & $+0.1\%$\\
\hline
    \end{tabular}
\end{center}
\caption{Predicted cross-sections 
at NNLO for 
$Z$ and a 120 GeV Higgs boson at the Tevatron 
and LHC.} 
\vspace{-0.3cm}
\label{cstablennlo}
\end{table}

The changes in predictions for cross sections at NNLO are seen in 
Table. \ref{cstablennlo}.  
Other than model dependence -- GMvarmod denotes the variation to 
${\tilde b} =2$ in the ${\cal O}(\alpha_S^3)$ term -- which reaches slightly more 
than $1\%$ for 14~TeV at the LHC, 
the maximum variations are of order $0.5\%$ at LHC. 
GMvarmod' is when the ${\cal O}(\alpha_S^3)$ terms fall with $Q^2$, and 
clearly exhibits a very small deviation.
Hence, as the changes in PDFs essentially guaranteed, there is far less 
uncertainty in predictions for cross sections due to uncertainties in
definitions of the GM-VFNS at NNLO than at NLO, exactly as we would expect 
for a theoretical ambiguity due only to the finite order at which one is 
working. The reduction in uncertainty is exactly as (usually) observed for
scale variations.

\begin{figure}
\vspace{-0.52cm}
\centerline{\hspace{-1.6cm}\includegraphics[width=0.45\textwidth]{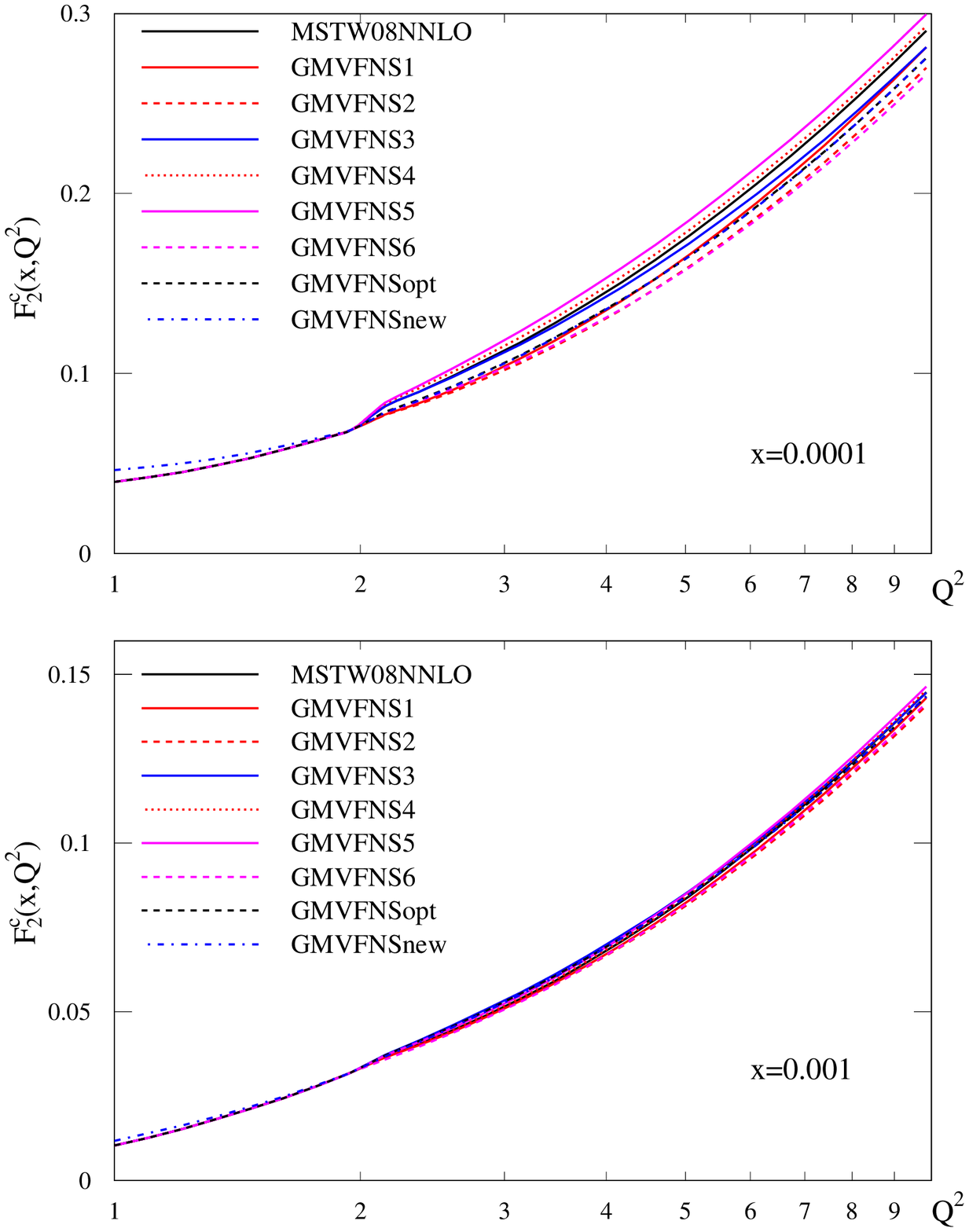}
\includegraphics[width=0.45\textwidth]{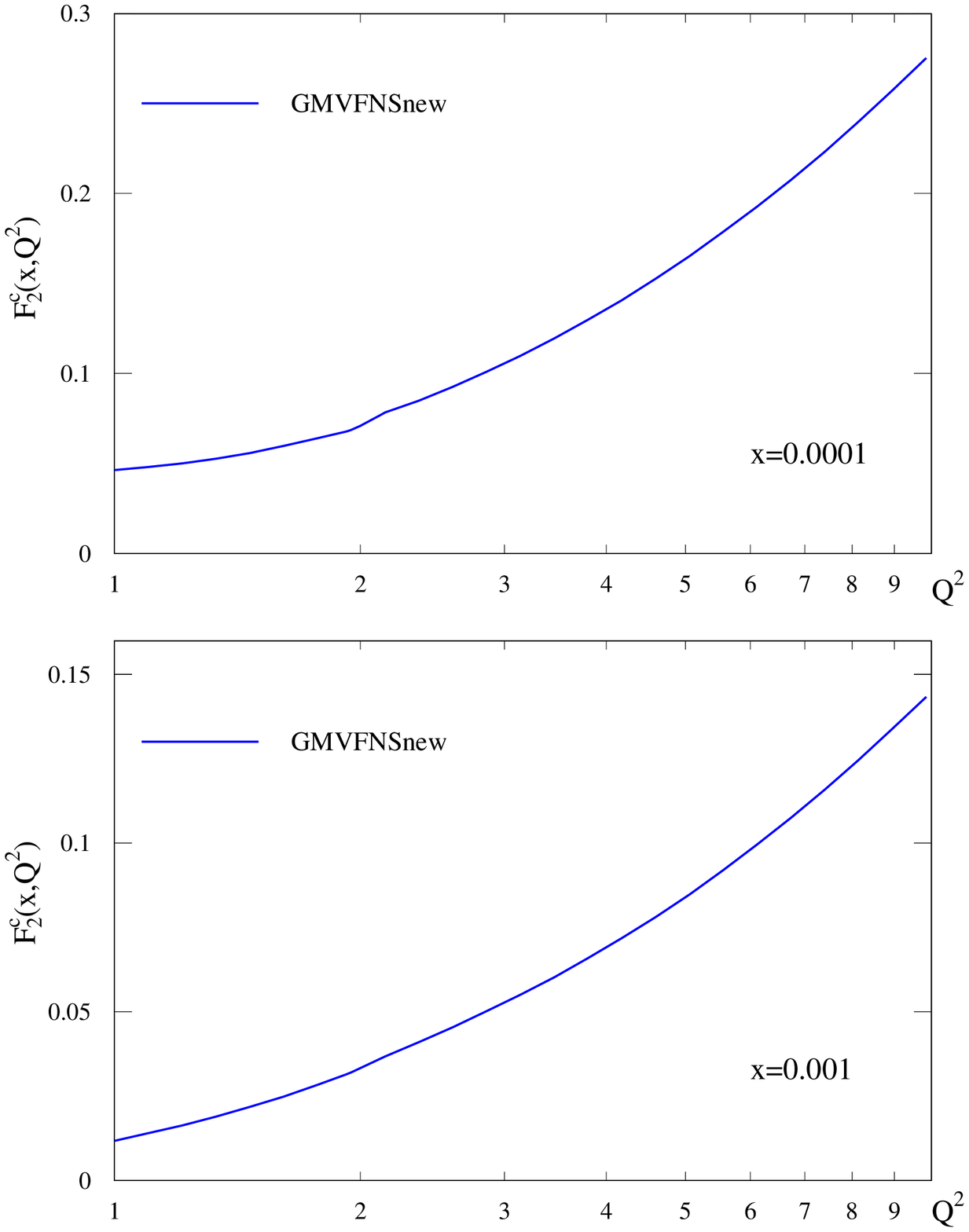}}
\vspace{-0.2cm}
\caption{The variation in $F_2^c(x,Q^2)$ generated from a variety of choices
of GM-VFNS at NNLO where the low $Q^2$ modelling has been modified in one,
improving the smoothness at the transition point (left-hand plots).  
In the right-hand plots only this choice is shown for clarity. }
\vspace{-0.5cm}
\label{gmvarf2cnew} 
\end{figure}

As a point of detail I note that at even at 
NNLO the smoothness across the transition point exhibited in the right of 
Fig. \ref{gmvarf2c} is not ideal. This is not so surprising at $x=10^{-4}$ 
since as noted a new $\ln(1/x)$ divergence in the evolution of the heavy 
quark is introduced into the quark evolution at 
NNLO which is not effectively cancelled by the subtraction terms.
However, at $x=0.001$ the slope immediately above the transition point is 
slightly flatter than below it. This is the same for all variations of the 
GM-VFNS, so in fact is not a feature of the definition of a GM-VFNS at all. 
It can be traced to the behaviour below $Q^2=m_h^2$, i.e. to the modelling of
the  ${\cal O}(\alpha_S^3)$ coefficient function. In the small-$x$ part of 
this in Eq. (\ref{sxmodel}) it is assumes that the $Q^2/m_h^2$ dependence of 
the constant term $\tilde b$ is the same as the known $\ln(1/x)$ term. 
Modifying this by a rather slowly varying factor of $(Q^2/m_h^2)^{0.1}$
results in the rather smoother behaviour shown in Fig. \ref{gmvarf2cnew}. 
Hence, this might be thought of as an improvement to this 
${\cal O}(\alpha_S^3)$ coefficient function. However, it affects the 
structure functions only below $Q^2=m_c^2$ for the charm contribution, i.e. 
below the MSTW2008 $Q^2$ cut, and the influence from the change in the 
very small amount of $F_2^b$ for $Q^2<m_b^2$ is very minor. Even without 
refitting the global fit changes by only about 1 unit, and so no significant 
change in PDFs occurs. More important, but very similar in practice, is an
improvement in threshold corrections which could be made in the same term.
Since the publication of the MSTW2008 PDFs more logarithms in the threshold 
correction have been calculated, see \cite{Presti:2010pd}. These can be 
included in the ${\cal O}(\alpha_S^3)$ coefficient function (though some 
approximation is needed since in \cite{Presti:2010pd} the value of one 
coefficient is presented only for specific $m_c^2/Q^2$ values, though the 
full result very recently appeared in \cite{Kawamura:2012cr}). The result is 
that this improvement itself makes the behaviour at the transition point 
smoother, and leads to a small modification of the structure function
both below, and at the transition point. The change at the transition point 
then persists as a constant if this term is frozen, though becomes relatively 
less important very rapidly, or quickly disappears if this term is defined 
to fall away like $m_h^2/Q^2$. Hence, even though this 
improvement will be included in future schemes, and is beneficial to the 
detailed behaviour of the structure function, within this definition of 
an NNLO GM-VFNS the effect on the PDFs is extremely small.     

Finally, it is an interesting question whether the changes in the GM-VFNS 
have any influence on the value of the quark masses preferred by the MSTW2008 
fits, as presented in \cite{Martin:2010db}. As noted in this previous study
there is little constraint on $m_b$ from fits, and indeed there is very 
little sensitivity to the GM-VFNS definition. However, there is a distinct 
effect on the conclusions regarding $m_c$ (see also \cite{H1ZEUS}). The default MSTW2008 
choice of charm mass
$m_c=1.4\GeV$ is chosen to be near the best fit value at NLO, which turns 
out to be $m_c=1.45\GeV$ with a very marginally better fit quality than
$m_c=1.4\GeV$. The same default mass is used in the studies in this article to
enable direct comparison to MSTW2008 results. However, when $m_c$ is allowed 
to vary using the optimal scheme at NLO the best fit value is for 
$m_c=1.32\GeV$, and the quality of the fit is 11 units better than for
$m_c=1.4\GeV$ and 17 units better than the best fit for $m_c=1.45\GeV$
using the default scheme. The fit does prefer the smoother behaviour near 
the transition point in the optimal scheme, but the slower turn on 
of $F_2^c$ at low $Q^2$ results in a preference for a lower mass. The variation
with scheme is much less at NNLO, and correspondingly the variation in 
the best fit mass is also smaller. It changes from $m_c=1.26\GeV$ in 
\cite{Martin:2010db} to $m_c=1.23\GeV$ using the optimal scheme. The 
best fit quality for the preferred mass is 7 units lower using the optimal
scheme at NNLO. Hence, the difference in the  values for the pole
charm mass extracted at NLO and NNLO has reduced from $0.19\GeV$ 
in \cite{Martin:2010db}, to 
$0.09\GeV$ using the optimum scheme, and both NLO and NNLO fits are within
their uncertainty for $m_c$ for an intermediate value of $m_c=1.28\GeV$.
Compatibility is certainly improved, but the values are rather low, a trend 
also seen for the charm mass defined in $\msb$ scheme extracted from
fits to structure function data in \cite{Alekhin:2010sv}. There is some 
evidence in this study that the convergence of the perturbation series in
the FFNS is quicker when using the $\msb$ scheme definition of the mass 
than when using the pole mass as is used in this study and is currently the 
most common framework. It would certainly be interesting to investigate this
in the context of GM-VFNS analyses. However, the situation is not quite 
the same in a GM-VFNS. The first dependence on scheme occurs for coefficient 
functions at order $\alpha_S^2$. Hence, for many GM-VFNS definitions, e.g. 
any ACOT variation and the default NNPDF choice, the mass-scheme dependence
would not set in until NNLO. For our definition the $\alpha_S$ coefficient 
functions only play a minor role at low $Q^2$, being either frozen at 
$Q^2=m_h^2$ or falling away quickly in the new optimal scheme. Hence, 
the mass-scheme dependence will be less influential at NLO in the GM-VFNS 
than in the FFNS, particularly in the optimal scheme. At NNLO the 
$\alpha_S^2$ coefficient functions are used at all scales, so the mass-scheme 
dependence will come into play fully. However, at low $Q^2$ where the 
mass-scheme dependence is undoubtedly most important there is the additional 
uncertainty due to the lack of the full ${\cal O}(\alpha_S^3)$ coefficient 
functions, so it is uncertain how much information the mass-scheme dependence 
can currently provide.

To summarise, I have considered variations in the definition of 
a GM-VFNS within the framework used in MSTW, and various other global fits 
for PDFs. In particular I have noted that none (in any framework) 
currently reduce to {\it exactly} the appropriate order FFNS expression at low 
$Q^2$ and also the appropriate order ZM-VFNS expression as $Q^2/m_h^2 
\to \infty$, though I have shown this can be easily achieved in the TR' scheme.
I have investigated the variation of both the heavy flavour 
structure function obtained and the PDFs extracted when 4 parameters 
controlling the definition of the GM-VFNS are varied within limits set by 
fit quality or sensible behaviour at NLO and NNLO. At NLO, variations in 
the resulting PDFs can be similar to the experimental PDF 
uncertainties at small $x$, and this is 
interpreted as a theoretical uncertainty. As such it is probably most 
appropriate to add this to the experimental PDF uncertainty linearly. 
However, as with other theoretical uncertainties it is impossible to 
gauge a confidence level. In principle a smooth GM-VFNS choice which also gives
the best fit should provide the central value. I would then consider the full 
variation  of GM-VFNS definitions in this article as somewhat conservative, and 
very likely to be more than a one sigma uncertainty.  
At NNLO there is much greater
stability, showing the success of the GM-VFNS approach, and the theoretical
uncertainty is much smaller than the experimental PDF uncertainty.  
One of these GM-VFNS choices is extremely smooth near the transition point,
a particular improvement at NLO, and reduces to {\it exactly} the correct 
asymptotic limits. It also gives a better fit quality, particularly
at NLO, and less variation in the preferred $m_c$ value between NLO and NNLO. 
This so-called ``optimal'' scheme seems to be a preferred option for 
global fits in future than the current default scheme used in the TR' 
definition, which was chosen originally simply for simplicity and 
continuity, and I recommend this new alternative for future global PDF fits.   

\vspace{-0.3cm}

\section*{Acknowledgements}

I would like to thank A. D. Martin, W. J. Stirling  and G. Watt for 
numerous discussions on PDFs, and S. Forte, P. Nadolsky and F. Olness 
for discussions 
on the topic of variable flavour number schemes. This work is
supported partly by the London Centre for Terauniverse Studies (LCTS), 
using funding
from the European Research Council via the Advanced Investigator Grant 267352.

\end{document}